\documentstyle[natbib,aas2pp4]{article}
\input{psfig}
\slugcomment{To appear in Physics Letters} \lefthead{I. Lopes and
J. Silk} \righthead{FROM GOLF AND SNO TO G: THE SUN AS A PROBE OF NEWTON'S CONSTANT}
\date{\today}
\begin{document}
\title{\bf  THE IMPLICATIONS FOR
HELIOSEISMOLOGY  OF EXPERIMENTAL UNCERTAINTIES IN NEWTON'S CONSTANT}
\author{IL\'IDIO P. LOPES$^{1,2}$\altaffilmark{\star} and JOSEPH SILK$^1$\altaffilmark{\star}}
\affil{$^1$Department of Physics, Denys Wilkinson Building,
Keble Road, Oxford OX1 3RH, United Kingdom \\
$^2$Instituto Superior T\'ecnico, Centro Multidisciplinar de
 Astrof\'\i sica, Av. Rovisco Pais, 1049-001 Lisboa, Portugal}
\altaffiltext{1}{Inquiries can be sent to {\bf
lopes@astro.ox.ac.uk} and {\bf silk@astro.ox.ac.uk}}
\setcounter{page}{1}
\newcommand{\RON}{read-out noise}
\newcommand{\beqa}{\begin{eqnarray}}\newcommand{\eeqa}{\end{eqnarray}}
\newcommand{\tnew}[1]{{\bf #1}}
%
\begin{abstract}
The  experimental uncertainties in G between different experiments have important implications for helioseismology.
We show that these uncertainties  for  the standard solar model  lead 
to a range in the 
value of the 
square of the sound speed in the nuclear region that is 
as much as  $0.15\%$ higher than the inverted helioseismic sound speed.
While a lower value of G is preferred
for the standard model, any definite prediction is masked by the uncertainties in the solar models available in the literature.
However future refinements of helioseismology 
with  an accuracy of the order of $10^{-3}$ to $10^{-4}$
in the square of the sound speed,
especially in combination with precision measurements of  the $^8B$ solar neutrino flux 
should be  capable of independently testing
these experimental values of G. 

\end{abstract}
\date{\today}

\twocolumn

\begin{table*}
\centerline{
\begin{tabular}{|l|l|l|}
\hline \hline
 Source & $G/(10^{-8} cm^3\;g^{-1}\;s^{-2} ) $ & Rel. Stand.   Uncert.   \\
  \hline
 Luo {\it et al.} (1999) & $6.6699(7)$ &$1.0\times 10^{-4}$    \\
 CODATA (1986) & $6.672 59(85)$ &$1.3\times 10^{-4}$   \\
 CODATA (1998) & $6.673 (10)$ &$1.5\times 10^{-3}$    \\
 Kleinevoss {\it et al.} (1999) & $6.67 35(29)$ &$4.3\times 10^{-4}$     \\
Gundlach \& Merkowitz  (2000) & $6.674 215 $ &$1.0\times 10^{-5}$   \\
Quinn {\it et al.} (2001) & $ 6.675 59(27) $ &$4.1\times 10^{-5}$  \\
Richman {\it et al.} (1999) & $6.683 11$ &$1.7\times 10^{-3}$ \\
Schwarz {\it et al.} (1999) & $6.6873(94) $ &$1.4\times 10^{-3}$  \\
Michaelis {\it et al.} (1996) & $6.715 40 (56)$ &$8.3\times
10^{-5}$ \\
\hline
\end{tabular}
}\caption{Summary of the recent experimental values of the
Newton's constant with their relative standard uncertainties. For
reference,  the values adopted by the CODATA
report in 1986 and 1998 are also represented.
Note - An extensive list of experiments
is present in the CODATA report, so we selected a subset of these
which represents the current determination of G. Details on
the experimental determination of G and on the different experiments
can be found in the Mohr \& Taylor (1999) report. }
\end{table*}

\begin{figure}
\centerline{\psfig{file=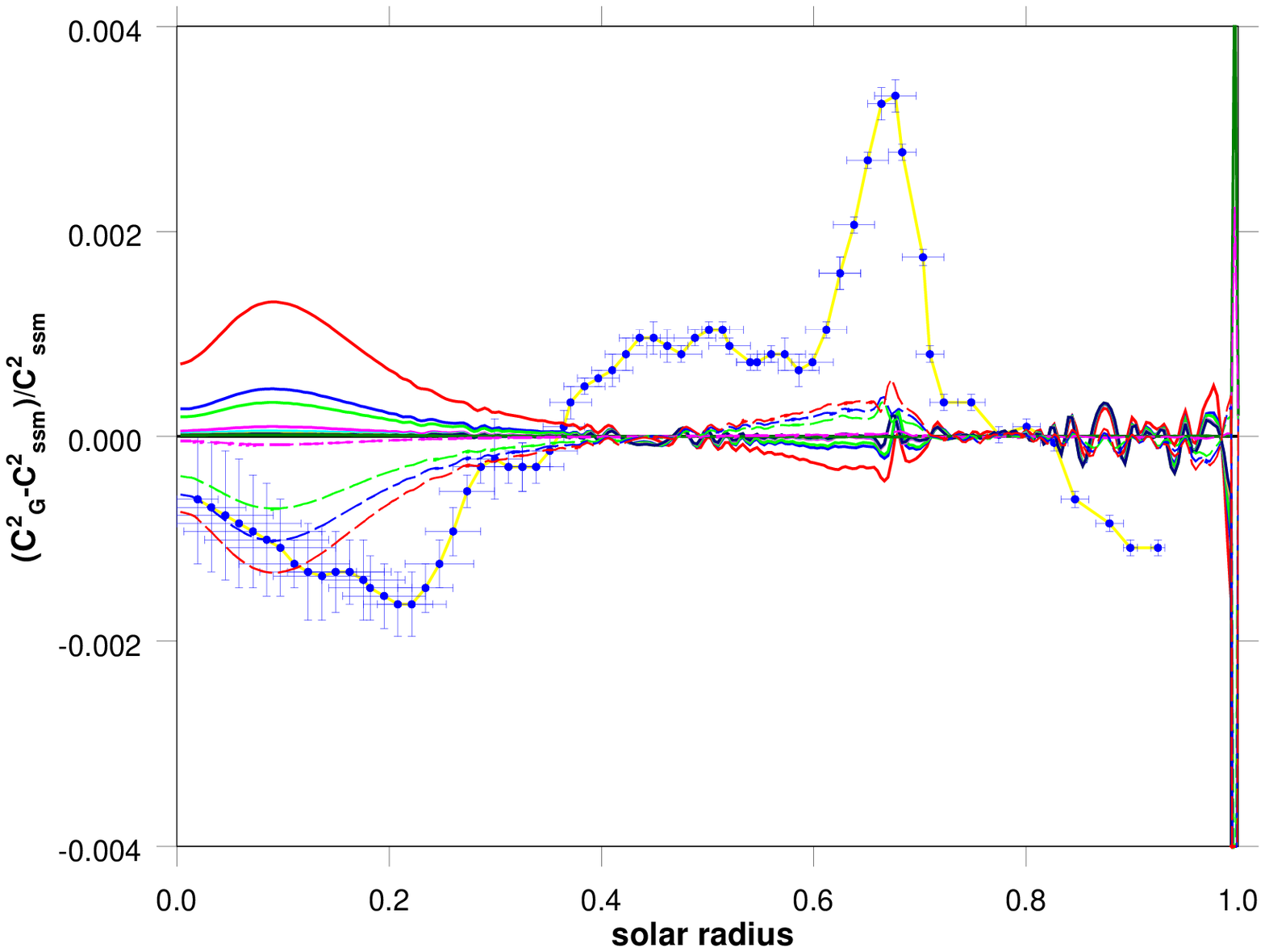,width=10.0cm,height=7.0cm}}
\vspace{0.0cm} \leftline{\hspace{0.5cm} \tiny \bf expanded scale}
\vspace{-1.5cm}
\centerline{\psfig{file=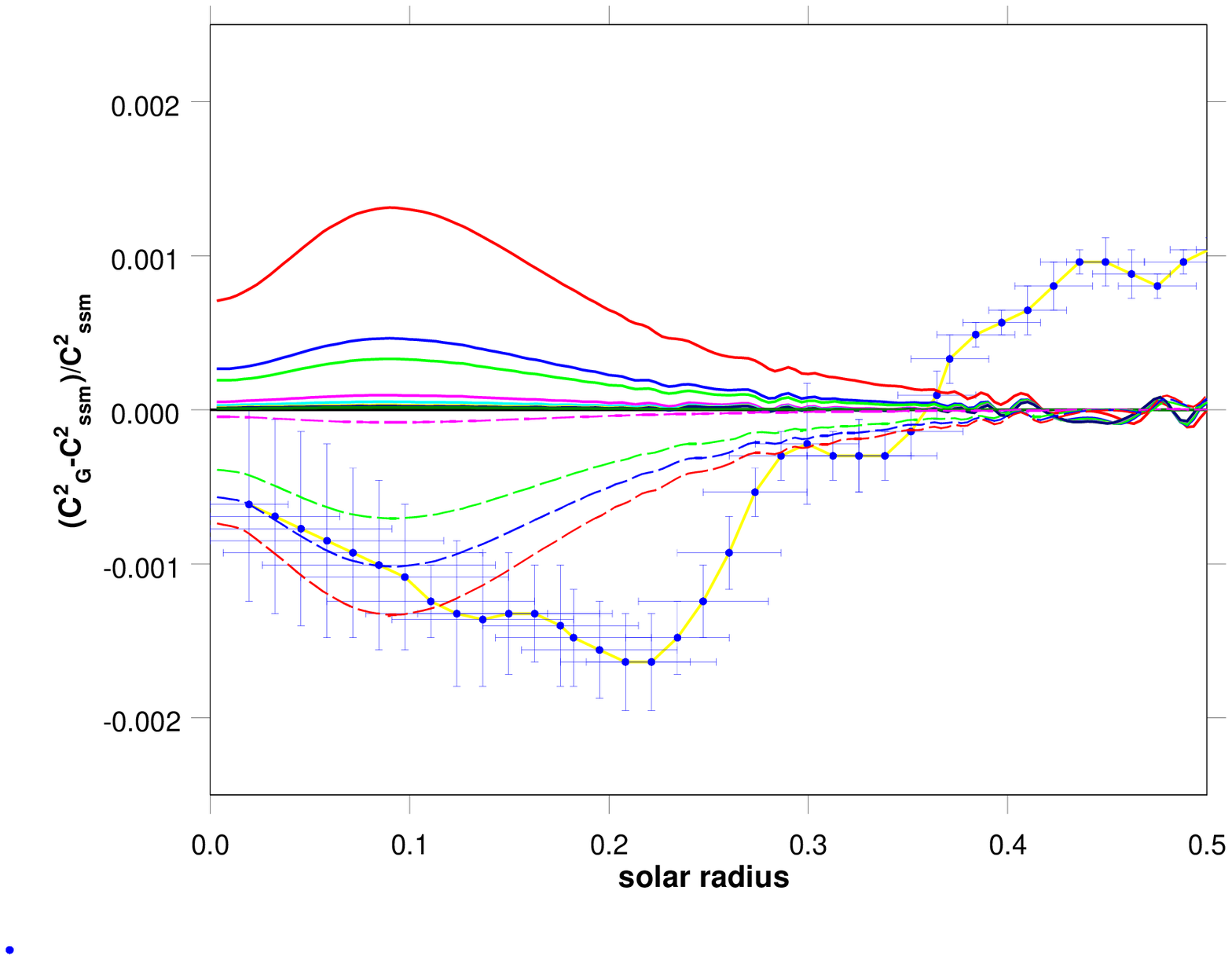,width=10.0cm,height=7.0cm}}
\caption{ \small The relative differences between the square of
the sound speed of the standard solar model and solar models with
different values of  Newton's gravitational constant (see Table 1).
The standard model used as reference corresponds to the value of
$G$ given in CODATA (1986). The continuous curves correspond to
models with $G$ values determined from different experiments:
Michaelis {\it et al.} (1996, red curve), Schwarz {\it et al.}
(1999, blue curve), Richman {\it et al.} (1999, green curve),
 Quinn {\it et al.} (2001, pink curve) and
Gundlach \& Merkowitz (2000, cyan curve)
The dashed curves correspond to  solar models computed with the
$G$ value determined by the Luo {\it et al.}(1999, pink curve) and
the values $G=6.65 \times 10^{-8} cm^3\;g^{-1}\;s^{-2}$ (green
curve),$6.64  \times 10^{-8} cm^3\;g^{-1}\;s^{-2}$ (blue curve)
and $6.63  \times 10^{-8} cm^3\;g^{-1}\;s^{-2}$ (red curve). The
yellow-blue curve with error bars represents the relative
differences between the squared sound speed in the Sun (as
inverted from solar seismic data) and a standard solar model
(Turck-Chi\'eze, Nghiem, Couvidat \& Turcotte 2001; Kosovichev
{\it et al.} 1999;1997). The horizontal bars show the spatial
resolution and the vertical bars are error estimates
\label{fig:1}}
\end{figure}
%
%

Newton's gravitational constant, G, stands apart from all the other
constants of physics in that the accepted uncertainty of a few per
thousand for G is several orders of magnitude larger than for
other fundamental constants (1998 CODATA report; Mohr \& Taylor
1999). An accurate determination of G is important for many fields
of modern physics, and  in particular for the new alternative
theories to general relativity that have started to emerge
(Forg\'acs \& Horv\'ath 1979; Albrecht \& Magueijo 1999; Barrow
1999; Barrow \& Magueijo 1998; Avelino, Martins, Rocha 2000;
Mbelek \& Lachi\`eze-Rey 2001). A common consequence of these
unified theories, applied to cosmology, is that they allow a space
and time dependence of the coupling constants, such as the speed
of light and  Newton's gravitational constants. Therefore an
accurate determination of G  in the laboratory is essential for
testing these new theories.

On the experimental side, the current interest in measuring G was
stimulated by a publication in 1996 by Michaelis, Haars \&
Augustin of a value of G that differed by $0.7\%$ from the
accepted value given in the previous 1986 CODATA report (see table
1; for recent reviews, see Quinn 2001, Mohr \& Taylor 1999). To
take this difference into account,
 the 1998 CODATA report recommends a value of G of
$6.673\times 10^{-8}\; \rm cm^3\; g^{-1}\; s^{-2}$ with an uncertainty
of $0.15\%$, some ten times worse than that in 1986. Whereas the
other fundamental constants are more accurately known than in
1986, the uncertainty in G has increased drastically. In an
attempt to improve the measurement of G, several groups around the
world have made  new measurements using a range of different
experimental methods (see table 1). The experimental  targets
represent an accuracy of between $0.01\%$ and $0.001\%$. New results have
recently been published, but in spite of the improved accuracy
obtained by the recent experiments, the disagreement between the
different measurements is still quite large (see table 1). In
particular, we refer to  the result of Luo {\it et al.} (1999),
which determines a value of G that is  $0.0026$  smaller that the
adopted value of CODATA in 1986. More recently, Gundlach \&
Merkowitz (2000) determined a value of G that is  $0.001215$ above
the 1998 CODATA value. Using two
independent methods Quinn {\it et al.} (2001)  found a value of G
$0.0026$ above the 1998
CODATA value. Even if the two more recent experiments lead to a
value above the 1998 CODATA report,
both the trend of other experiments (Mohr \& Taylor 1999),
and the values of Gundlach \& Merkowitz (2000)  and Quinn {\it et al.}
(2001), do not seem to agree (see table 1). The accuracy of the
experiments has improved by as much as $1\;10^{-5}$, but the
disagreement between the different results is of the order of
$1.4\;10^{-3}$, which is quite striking. In general the
experimental values of G varies between $6.669\; \rm 10^{-8} cm^3
g^{-1} s^{-2}$ and $6.715\; 10^{-8} \rm cm^3 g^{-1} s^{-2}$.


In this Letter we study the consequences of these new measurements
 of G  for the evolution of the Sun. We use
the solar seismic data and the $^8B$ neutrino flux as probes.
Furthermore, we confront these results with the recent solar
neutrino measurements  of Super-Kamiokande (SK; Fukuda {\it et
al.} 2001) and the Sudbury Neutrino Observatory (SNO; Ahmad {\it
et al.} 2001).

\begin{figure}[t]
\centerline{\psfig{file=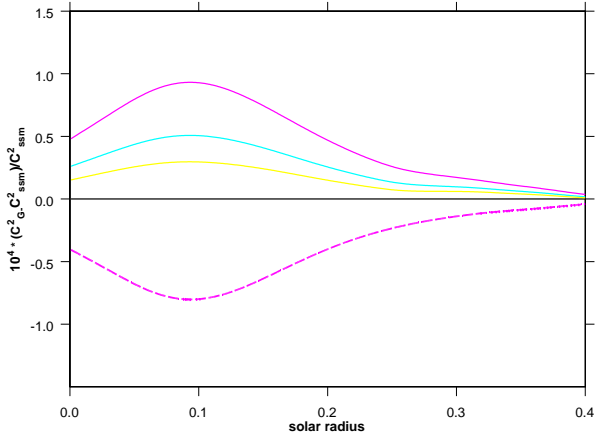,width=9.0cm,height=7.0cm}}
\caption{ \small The relative differences between the square of
the sound speed of the standard solar model and solar models with
different values of Newton's gravitational constant (see Fig.~1
and  Table 1). The continuous curves correspond to models with $G$
values determined from different experiments: Quinn {\it et al.}
(2001, pink curve), Gundlach \& Merkowitz (2000, cyan curve) and
Kleinevoss {\it et al.} (1999, yellow curve). The pink dashed
curve corresponds to the value of solar models with the $G$ value
determined by the Luo {\it et al.} (1999). Note - The vertical
axis in this figure is multiplied by $10^4$. \label{fig:2}}
\end{figure}

The idea of using stellar evolution to constrain the possible
value of G was  originally proposed by Teller (1948), who stressed
that the evolution of a star was strongly dependent on G. The
luminosity of a main sequence star can be expressed as a function
of  Newton's gravitational constant and its mass by using homology
relations (Teller 1948, Gamow 1967, Kippernhahn \& Weigert 1994).
In the particular case that the opacity is dominated by free-free
transitions, Gamow (1967) found that the luminosity of the star is
given approximately by $L\approx G^{7.8}\;M^{5.5}$. In the case of
the Sun, this would mean that for higher values of G, the burning
of hydrogen will be more efficient and the star evolves more
rapidly, therefore we need to increase the initial content of
hydrogen to obtain the present observed Sun. In a numerical test
of the previous expression, Delg'Innocenti {\it et al.} (1996)
found that  low-mass stars evolving from the Zero Age Main
Sequence to the red giant branch satisfy
 $L\propto G^{5.6}\;M^{4.7}$, which agrees
to within $10\%$ of the numerical results, following the idea that
Thomson scattering contributes significantly to the opacity inside
such stars. Indeed, in the case of the opacity being dominated by
pure Thomson scattering, the luminosity of the star is given by
$L\approx G^4\;M^{3}$. It follows from the previous analysis that
the evolution of the star on the main sequence is  highly
sensitive to the value of G. Following this idea, several attempts
to directly check the sensitivity of G to stellar evolution, and
in particular its temporal variation, have been previously
performed. The effect of a possible
 time-dependence of G on luminosity has been
studied  in the case of globular cluster H-R diagrams but has not
yielded any stronger constraints than those relying on celestial
mechanics (Will 1993, reference therein). In 1998, Guenther and
collaborators used solar acoustic oscillation spectra available at
that time to constrain the time variation of G, setting an upper
limit on the variation of G that was  $1.6\times 10^{-12}
yr^{-1}$, almost one order of magnitude smaller than the
constraints obtained by binary pulsar timing measurements (Will
1993). In this context, the evolution of main sequence stars like
the Sun presents an excellent probe for discussing new
experimental values of G. This argument is validated by a strong
constraint that can be used to diagnose the internal structure of
our star, namely in the nuclear region,  through the new results of helioseismology.
\begin{figure}[t]
\centerline{\hspace{0.3cm}\psfig{file=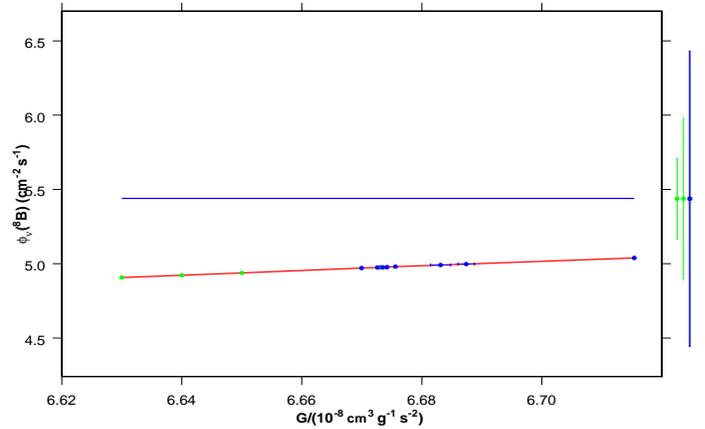,width=10.0cm,height=7.0cm}}
\caption{\small This figure shows the variation of the predicted neutrino flux
of $^8B$ with the variation of Newton's gravitation
constant. The red curve corresponds to a  solar model with
standard physics, the only difference being the fact that we
 let the value of Newton's gravitational constant  vary continuously. The blue points with error bars correspond to the following experimental values (from the left to the right) of all the models listed in table 1. The green points correspond to a $^8B$ neutrino flux computed for solar models with the value of Newton's constant $G=6.63,6.64,6.65$ in units of $ 10^{-8} \rm cm^3\;g^{-1}\;s^{-2}$.
The vertical blue line at the right of the figure gives us the measured value of the combined SNO-SK data with an error bar of $18\%$
(Ahmad {\it et al.} 2001). The sequence of the other two vertical green lines corresponds to accuracy targets of SNO that may be feasible in the future (respectively 10\% and 5\%, Wark 2001, private communication).
\label{fig:3}}
\end{figure}
A larger value of  Newton's gravitational constant increases
the gravitational force, which, for stars on the main sequence in hydrostatic equilibrium, is compensated
by an increase in the rate of
thermonuclear reactions. This leads to an increase of the central
temperature and has two main consequences: since central pressure
support must be maintained, the central density is increased in
the solar models with G larger than the reference model (in our
case CODATA 1986), and since more hydrogen is burnt at the centre
of the Sun, the central helium abundance and the central molecular
weight are larger than in standard solar models.

The Sun is a unique star for research because its proximity allows a superb
quality of solar data, enabling precision measures of its luminosity, mass,
radius and chemical composition.  Therefore it naturally becomes a privileged
tool to be used as a laboratory for physics.  In recent years, different
groups around the world have produced solar models in the framework of
classical stellar evolution, taking into account the best known physics as
well as all the available observational seismic data. This has led to the
determination of a well-established model for the Sun, the so-called standard
solar model (Turck-Chi\'eze \& Lopes 1993; Christensen-Dalsgaard {\it et al.}
1996; Brun, Turck-Chi\`eze \& Zahn 1999; Provost, Berthomieu \& Morel 2000;
Turck-Chi\'eze, Nghiem, Couvidat \& Turcotte 2001, Turck-Chi\`ze S. {\it et
al.} 2001a; Bahcall, Pinsonneault, Basu 2001), for which the acoustic modes
are in very good agreement with observation. Furthermore, this model has
established considerable consensus among the different research groups,
concerning the predictions of the solar neutrino fluxes, and has
unambiguously helped define the difference between the theoretical
predictions and the experimental results. In this context, we use the
standard solar model as a reference to test the new experimental measurements
of $G$ (see Table 1). We have produced different solar models which are
distinguished from the solar standard model by adopting different values of
G. As usual, the model starts to evolve from a standard primordial chemical
composition star to reach the present Sun with the observed luminosity and
radius at its present age 4.6 Gyr, by readjusting the initial helium
abundance and the mixing length parameter (Turck-Chi\`eze {\it et al.} 2001;
Lopes, Silk and Hansen 2001; Lopes, Bertone \& Silk 2001).  There has been
considerable discussion recently regarding the precise value of the solar
radius and solar luminosity, therefore in order to obtain the present solar
models for different values of $G$ we have performed a calibration by
choosing the value of mixing-length and the initial content of helium in such
a way that the present luminosity and the present solar radius are reproduced
with an accuracy better than $10^{-7}$(Lopes \& Silk 2002, in
preparation).  This calibration precision is much higher than the precision
of the global quantities which is of the order of a few thousand.
At this stage, it is worth noticing  that the standard evolution of the Sun is
independent of the total mass of the star. The observational determination
of the product $GM_\odot$, i.e., the product of Newton's constant with the
total mass of the Sun, the so-called Gaussian constant, is known
with an accuracy better than $10^{-7}$, therefore the $GM_\odot$ product
can be used to explicitly write the equations of stellar structure as a
function of Newton's constant.
Actually, it is this fact that provides us with a possible means of
probing  the value of G by using the evolution of the Sun and the highly
accurate results of helioseismology and solar neutrinos (Lopes \& Silk
2002, in preparation).

In Fig.~1, we compare the square of the sound speed for different solar models with the
new experimental values of $G$ and the solar standard model. In the same figure,
we show the square of the sound-speed as inferred for the present Sun by using the
data from Global Oscillations at Low Frequency (GOLF; Gabriel {\it et al.}
1995) and Michelson Doppler Imager (MDI; Scherrer {\it et al.}
1995) experiments. It follows from our analysis that the new experimental
values of G determined by Michaelis {\it et al.} (1996), Schwarz {\it et al.}
  (1999), Richman {\it et al.} (1999), produce changes in the profile
  of the sound speed, compared with the inverted sound speed, that
  are larger than the differences currently obtained with the solar
  standard model. Conversely, models computed with lower values of $G$,
  of the order of $6.63\; 10^{-8} \rm cm^3 g^{-1} s^{-2}$ reproduce the
  differences observed
between the solar standard model and the inverted sound speed from
the more recent seismic data. Naturally, the inversion of the
sound speed in the center is not totally reliable but we infer
that the seismology seems to favor a lower value of G. If
we consider that among  the present values of $G$ the more reliable
 are those
of Quinn {\it et al.} (2001) and Gundlach \& Merkowitz (2000),
than we anticipate
that in the coming years solar physics and seismology
can progress together to a level such that the accuracy of the square of the
 sound speed can be obtained with an accuracy of 1 part in $10^4$.
In this  case it follows
follows that the standard solar model will be
capable of distinguishing between the more likely values of G (see
Fig.~2).

Indeed, it is important to remark that the inversion of the sound
speed is still uncertain in the central region due to the lack of
seismic data, mainly due to the small number of acoustic modes
that reach the nuclear-burning region. The inversions are not very
reliable at the surface, above $98 \%$ of the solar radius, due to
a poor description of the interaction of acoustic waves with the
radiation field and  turbulent convection, namely, in the
superadiabatic region (Lopes \& Gough 2001).  In spite of these
uncertainties, it is not possible to explain the large differences
in the nuclear region obtained by these experimental values of G
in the solar standard model. Indeed, a positive difference of as
larger as  $0.3\%$ cannot be accommodated by our  present
understanding of the internal structure of the Sun. However, a
lower value of G, typically of the order of $6.63\; 10^{-8} \rm cm^3
g^{-1} s^{-2}$, produces changes in the structure comparable to the
helioseismic sound speed and accurately reproduces its shape in
the nuclear region.

If we believe in the diagnostic capability of the seismic
techniques presented here, a lower value of G is better
accommodated in the present picture of the evolution of the Sun
than  the experimental values of $G$ measured by Quinn {\it et al.}
(2001) and Gundlach \& Merkowitz (2000), among others. However,
we stress that in order to determine with certainty the impact of
the new values of $G$ on the evolution of the Sun, a more careful
analysis of this problem must be made.


The SNO collaboration have  published their result for the $^8B$ flux measured by
neutrino-electron scattering reactions and reported a lower $^8B$
flux as compared to the theoretical predictions of the standard
solar model (Bahcall, Pinsonneault and Basu 2001; Turck-Chi\`eze,
Nghiem, Couvidat \& Turcotte 2001). Furthermore, this result is in
agreement with the $^8B$ flux measured by Super-Kamiokande
detector through the same reaction. Therefore, the experiment
confirms the deficit of solar neutrino fluxes of the Chlorine
experiment of Davis {\it et al.} (1998) and subsequently confirmed
by Kamiokande, and by the Gallium experiments SAGE (Abdurashitov
{\it et al.} 1996), GALLEX  (Kirsten {\it et al.} 2000) and GNO(
Belloti {\it et al.} 2000).

The production of $^8B$ takes place in the inner $2\%$ of the
solar mass core. The $^8B$ decay reaction presents the strongest
dependence on the temperature: the $^8B$ neutrino production
is maximum at quite small radii, $5\%$ 
of the solar radius, and its generation is confined to  the region
between $2\%$ and  $7\% $ of the solar radius.  Consequently, this
flux of neutrinos becomes the best signature of the temperature at
the center of the Sun. Indeed, if the SNO measurement of the $^8B$
neutrino fluxes is correct, the central temperature of the
standard solar model is within less than $0.5\%$ of the
temperature deduced from the measured $^8B$ neutrino flux (Bahcall
2001, Turck-Chi\`eze 2001). It follows that the high constraint on
the solar central temperature imposed by the SNO results can be
used to constrain some physical processes occurring in the center
of the Sun, or even test some SUSY dark matter particles (Lopes \&
Silk 2002; Lopes Hansen \& Silk 2002; Lopes, Bertone \& Silk 2002).

The evolution of the Sun on the main sequence occurs  under
hydrostatic equilibrium, and accordingly the kinetic energy of
electrons and nuclei in the Sun is proportional to the
gravitational potential. Therefore the central temperature can be
used to measure G. In Fig.~2 we present a solar model for
different values of G. At present the error bar of SNO does not
allow us to identify which is the best value of G. Nevertheless, SNO is
expected to attain an accuracy of $10\%$ or even $5\%$ in
future years (Wark 2001, private communication).
With this precision
it is not possible to unequivocally determine the best value of G,
based solely upon  the constraint imposed by SNO neutrinos.
However   these results, combined
with other neutrino experiments and  with a better
understanding of the mechanisms of neutrino oscillations,
will enable us to open
a new means of  constraining the value of G in the solar interior. The
information provided by the neutrino experiments is quite
significant because it constitutes  an independent test of G
complementary to the one provided by helioseismology.

It has been known for  the last two hundred years that Newton's
constant is very difficult to measure accurately. Simply stated, G
is determined by measuring the gravitational attractive force
between two masses at a known distance apart. The problem is that
the gravitational attraction between two laboratory-sized masses
is simply too small. However, using a very large body like the Sun
and the solar acoustic spectrum, it is possible to constrain the
gravitational self-attraction and, in so doing, test the new
experimental values of $G$. It follows from our analysis that the
low values of G seem to be favoured. However, the solar standard
model is  quite complex, and the accurate determination of
G can be masked by other uncertainties in the solar model. Only a
systematic study of these uncertainties can lead to an accurate
determination of G. However, the fact that the neutrino flux
measured by SNO (and possibly other neutrino experiments) can be
combined with a better model for the oscillation properties of neutrinos
will provide  a promising means of potentially determining G with   improved
accuracy, specially if current experimental error bars are significantly
reduced.

IPL is grateful for support by a grant from Funda\c c\~ao para a
Ci\^encia e Tecnologia. The authors thanks S. Turck-Chi\`eze for
stimulating discussions. It is a pleasure to thank also
H. M. Antia, S. Basu, J. Christensen-Dalsgaard,
M. Lachi\`eze-Rey, A. Lynas-Gray, J. P. Mbelek, and H. Meyer 
for informative comments.

\bibliography{helioseismology}




\end{document}